# Gravitation as a Casimir interaction

Bo E. Sernelius

*Theory and Modeling, Department of Physics, Chemistry and Biology, Linköping University, SE-581 83 Linköping, Sweden.*

**Gravitation is considered to be one of the four fundamental interactions in nature. However, one has so far failed to observe the graviton, the quantum particle that is believed to transmit the gravitational force at a distance – the analogue to the photon in electromagnetism. Maybe it is now time to re-evaluate the status of the gravitation as a fundamental interaction. Here, we propose a completely new interpretation of gravitation. In this description the gravitational force is no longer a fundamental force. It is an induced force, a dispersion force, and the analogue to the Casimir force[1] in electromagnetism. The fundamental force is in our description a force between particles with a parabolic interaction potential. In our model the nucleons are made up from these particles. We find the retarded dispersion force[2] between these composite particles has the correct distance dependence, $1/r$. If this interpretation is correct it has a broad range of implications. Our view on the fundamental concept mass is altered; our view on the expansion of the Universe may change.**

The objective of this work is to find out if a more fundamental interaction can give rise to the gravitational force. We believe that a candidate for a fundamental force field should have closed classical orbits. This belief is based on the following hand-waving arguments: In the classical picture the particle goes around and around in its orbit; in the semi-classical picture a wave function goes around and around along the orbit and is superposed on itself; if the orbit is closed and the energy is right the wave interferes constructively and there is a self-sustained solution to the Schrödinger equation; if not the wave function will disintegrate and the state will not survive. There are only two central force fields where all bound orbits are closed: one has the interaction potential $V \sim r^{-1}$ and the other has $V \sim r^2$. Examples of the first type are the Coulomb potential and the here challenged gravitational potential. The second type is known as a harmonic oscillator potential and is our obvious choice of candidate. Our notation throughout this letter is in analogy with the electromagnetic case. We put a tilde above the quantities to distinguish them from the electromagnetic counterparts. We assume that the particles have charge $\pm \tilde{q}$ and postulate that the electric field from a charge $\tilde{q}$ is $\tilde{\mathbf{E}} = \tilde{q}\mathbf{r}$. This gives the scalar potential the form $\tilde{\varphi}(r) = -\tilde{q}r^2/2$. This type of interaction leads to composite particles that are neutral and have no direct interaction; dispersion interactions due to fluctuations are possible. The particles can never be found individually; they are always in pairs. It takes an infinite energy to separate them completely. If we assume that they are each other's anti-particles the pair will turn into two pairs when one tries to separate them past a critical distance. Let us clarify the model a little bit more: These fundamental particles are not replacing the quarks in the nucleons; they are a complement. Also the electron, which is considered to be a point particle, can be associated with a cloud of these fundamental particles. The energy density in the electric field of the electron diverges at the position of the electron and this can lead to the creation of particle-antiparticle pairs.

To find the effect from fluctuating electric dipoles we have to start with the field from a dipole, $\tilde{\mathbf{p}} = \tilde{q}\mathbf{d}$. It is found to be $\tilde{\mathbf{E}}(\mathbf{r}) = -\tilde{\mathbf{p}}$. Thus the field from a dipole is just minus the dipole itself. There are no quadrupole or higher order multi-pole contributions as opposed to in the ordinary electromagnetic theory. Furthermore the field has no spatial dependence. This holds for any static distribution of charges within the composite particle – the field lacks spatial dependence. This implicates that the only static field is a dipole field and this dipole field is constant throughout all space, independent of the position of the particles. There are no other multi-pole fields. This is very encouraging. In the ordinary electromagnetic theory the dispersion potential between two atoms from dipole-dipole interactions vary as $-r^{-6}$ in the non-retarded van der Waals range and as $-r^{-7}$ in the retarded Casimir limit. However for smaller separations higher order multi-pole contributions enter and complicate the separation dependence; this prevents the dispersion interaction in electromagnetism to be mistaken for a fundamental interaction.

From Einstein's two postulates in special relativity follows that there has to be a magnetic field companion to the electric field. Furthermore these two fields obey the two homogeneous Maxwell's equations, $\nabla \times \tilde{\mathbf{E}} + \frac{1}{c}\partial \tilde{\mathbf{B}}/\partial t = 0$ and $\nabla \cdot \tilde{\mathbf{B}} = 0$. This means that we may introduce scalar and vector potentials, $\tilde{\varphi}$ and $\tilde{\mathbf{A}}$, respectively, where $\tilde{\mathbf{B}} = \nabla \times \tilde{\mathbf{A}}$ and $\tilde{\mathbf{E}} = -\nabla \tilde{\varphi} - \frac{1}{c}\partial \tilde{\mathbf{A}}/\partial t$, all in complete analogy with in electromagnetism. With the retarded potentials,

$$\tilde{\varphi}(\mathbf{r},t) = -\frac{1}{2}\int d^3 r' \tilde{\rho}(\mathbf{r}',t-R/c)R^2;$$
$$\tilde{\mathbf{A}}(\mathbf{r},t) = \frac{1}{2c}\int d^3 r' \tilde{\mathbf{J}}(\mathbf{r}',t-R/c)R^2; \quad \mathbf{R} = \mathbf{r} - \mathbf{r}',$$
(1)

where $\tilde{\rho}$ and $\tilde{\mathbf{J}}$ are the charge and current densities, respectively, we find the fields from a time dependent dipole. We perform the derivation along the lines used by Heald and Marion[3] on a Hertzian dipole and arrive at

$$\tilde{\mathbf{E}}(\mathbf{r},t) = \left\{-[\tilde{p}] + 2[\dot{\tilde{p}}]\tfrac{r}{c} - [\ddot{\tilde{p}}]\left(\tfrac{r}{c}\right)^2\right\}\cos\theta \hat{r} + \left\{[\tilde{p}] - \tfrac{1}{2}\left(\tfrac{r}{c}\right)[\dot{\tilde{p}}] + \tfrac{1}{2}[\ddot{\tilde{p}}]\left(\tfrac{r}{c}\right)^2\right\}\sin\theta \hat{\theta};$$
$$\tilde{\mathbf{B}}(\mathbf{r},t) = \left\{-\left(\tfrac{r}{c}\right)[\dot{\tilde{p}}] + \tfrac{1}{2}\left(\tfrac{r}{c}\right)^2[\ddot{\tilde{p}}]\right\}\sin\theta \hat{\varphi},$$
(2)

where a dot means the time derivative and square brackets that the function within the brackets is determined at retarded times, $t - R/c$. We have chosen a spherical coordinate system with its *z*-axis along **p**.

We define the polarizability, $\tilde{\alpha}$, for a composite particle through: $\tilde{\mathbf{p}} = \tilde{\alpha}\tilde{\mathbf{E}}$. The interaction between two particles, 1 and 2, is found from realizing that a polarization of particle 1 gives rise to an electric field at the position of particle 2. This field polarizes particle 2, which results in a field at the position of particle 1. Closing this loop results in self-sustained fields – normal modes. We find these in analogy with the treatment of the two-atom system in the book: *Surface Modes in Physics*[2]: We may reformulate the electric field as

$$\tilde{\mathbf{E}}(\mathbf{r},t) = (\hat{p}\cdot\hat{r})\hat{r}\left[-\tilde{p}(t-r/c) + 2\left(\tfrac{r}{c}\right)\dot{\tilde{p}}(t-r/c) - \left(\tfrac{r}{c}\right)^2 \ddot{\tilde{p}}(t-r/c)\right] \\ + \left[\hat{p}-(\hat{p}\cdot\hat{r})\hat{r}\right]\left[-\tilde{p}(t-r/c) + \tfrac{1}{2}\left(\tfrac{r}{c}\right)\dot{\tilde{p}}(t-r/c) - \tfrac{1}{2}\left(\tfrac{r}{c}\right)^2 \ddot{\tilde{p}}(t-r/c)\right], \quad (3)$$

and its Fourier transform with respect to time becomes

$$\tilde{\mathbf{E}}(\mathbf{r},\omega) = \tilde{p}(\omega)e^{i\omega r/c}\left\{(\hat{p}\cdot\hat{r})\hat{r}\left[-1 + 2(-i\omega r/c) + (\omega r/c)^2\right] \\ + \left[\hat{p}-(\hat{p}\cdot\hat{r})\hat{r}\right]\left[-1 + \tfrac{1}{2}(-i\omega r/c) + \tfrac{1}{2}(\omega r/c)^2\right]\right\}. \quad (4)$$

On tensor form we may write

$$\tilde{\mathbf{E}}(\mathbf{r},\omega) = e^{i\omega r/c}\left\{\tilde{\tilde{\alpha}}\left[-1 + 2(-i\omega r/c) + (\omega r/c)^2\right] + \tilde{\tilde{\beta}}\left[-1 + \tfrac{1}{2}(-i\omega r/c) + \tfrac{1}{2}(\omega r/c)^2\right]\right\}\tilde{\mathbf{p}}(\omega), \quad (5)$$

where the tensors are: $\tilde{\tilde{\alpha}} = r_\mu r_\nu / r^2$; $\tilde{\tilde{\beta}} = \delta_{\mu\nu} - r_\mu r_\nu / r^2 = \tilde{\tilde{I}} - \tilde{\tilde{\alpha}}$. We let all tensors have double tildes to distinguish them from our fields. Things become very simple if we choose the third principle axis to point along **r**. Then we have

$$\tilde{\tilde{\alpha}} = \begin{pmatrix} 0 & 0 & 0 \\ 0 & 0 & 0 \\ 0 & 0 & 1 \end{pmatrix}; \quad \tilde{\tilde{\beta}} = \begin{pmatrix} 1 & 0 & 0 \\ 0 & 1 & 0 \\ 0 & 0 & 0 \end{pmatrix}, \quad (6)$$

and $\tilde{\tilde{\alpha}}^2 = \tilde{\tilde{\alpha}}$; $\tilde{\tilde{\beta}}^2 = \tilde{\tilde{\beta}}$; $\tilde{\tilde{\alpha}}\cdot\tilde{\tilde{\beta}} = 0$.

If we now let

$$\tilde{\tilde{\gamma}} = e^{i\omega r/c}\left\{\tilde{\tilde{\alpha}}\left[-1+2(-i\omega r/c)+(\omega r/c)^2\right]+\tilde{\tilde{\beta}}\left[-1+\tfrac{1}{2}(-i\omega r/c)+\tfrac{1}{2}(\omega r/c)^2\right]\right\}, \qquad (7)$$

we may set up the self-sustained fields between two polarizable particles as

$$\begin{aligned}\tilde{\mathbf{p}}_2(\omega) &= \tilde{\alpha}_2(\omega)\tilde{\mathbf{E}}_1(\mathbf{r}_2,\omega) = \tilde{\alpha}_2(\omega)\tilde{\tilde{\gamma}}\tilde{\mathbf{p}}_1(\omega); \\ \tilde{\mathbf{p}}_1(\omega) &= \tilde{\alpha}_1(\omega)\tilde{\mathbf{E}}_2(\mathbf{r}_1,\omega) = \tilde{\alpha}_1(\omega)\tilde{\tilde{\gamma}}\tilde{\mathbf{p}}_2(\omega),\end{aligned} \qquad (8)$$

where $\tilde{\mathbf{E}}_i(\mathbf{r}_j,\omega)$ is the electric field at the position of particle *j* caused by the polarized particle *i*. Eliminating $\mathbf{p}_2$ gives

$$\tilde{\mathbf{p}}_1(\omega) = \tilde{\alpha}_1(\omega)\tilde{\tilde{\gamma}}\tilde{\alpha}_2(\omega)\tilde{\tilde{\gamma}}\tilde{\mathbf{p}}_1(\omega) = \tilde{\alpha}_1(\omega)\tilde{\alpha}_2(\omega)\tilde{\tilde{\gamma}}^2\tilde{\mathbf{p}}_1(\omega), \qquad (9)$$

where in the last step we have assumed that the polarizabilities are isotropic. Now, we have

$$\tilde{\tilde{A}}\tilde{\mathbf{p}}_1(\omega) = \left(\tilde{\tilde{I}} - \tilde{\alpha}_1(\omega)\tilde{\alpha}_2(\omega)\tilde{\tilde{\gamma}}^2\right)\tilde{\mathbf{p}}_1(\omega) = 0. \qquad (10)$$

This system of equations has non-trivial solutions, the normal modes, if $\left|\tilde{\tilde{A}}\right| = 0$. Now,

$$\tilde{\tilde{\gamma}}^2 = e^{i2\omega r/c}\left\{\tilde{\tilde{\alpha}}\left[-1+2(-i\omega r/c)+(\omega r/c)^2\right]^2+\tilde{\tilde{\beta}}\left[-1+\tfrac{1}{2}(-i\omega r/c)+\tfrac{1}{2}(\omega r/c)^2\right]^2\right\}, \qquad (11)$$

and

$$\begin{aligned}\left|\tilde{\tilde{A}}(\omega)\right| &= \left[1 - \tilde{\alpha}_1(\omega)\tilde{\alpha}_2(\omega)e^{i2\omega r/c}\left[-1+2(-i\omega r/c)+(\omega r/c)^2\right]^2\right] \\ &\times \left[1 - \tilde{\alpha}_1(\omega)\tilde{\alpha}_2(\omega)e^{i2\omega r/c}\left[-1+\tfrac{1}{2}(-i\omega r/c)+\tfrac{1}{2}(\omega r/c)^2\right]^2\right]^2.\end{aligned} \qquad (12)$$

We will need the results on the imaginary frequency axis,

$$\left|\tilde{\tilde{A}}(i\omega)\right| = \left[1 - \tilde{\alpha}_1(i\omega)\tilde{\alpha}_2(i\omega)e^{-2|\omega|r/c}\left[-1 + 2(|\omega|r/c) - (\omega r/c)^2\right]^2\right] \\ \times \left[1 - \tilde{\alpha}_1(i\omega)\tilde{\alpha}_2(i\omega)e^{-2|\omega|r/c}\left[-1 + \tfrac{1}{2}(|\omega|r/c) - \tfrac{1}{2}(\omega r/c)^2\right]^2\right]^2. \tag{13}$$

Now, the zero-point energy for the system of two interacting particles, $r$ apart, relative the energy when they are at infinite separation is

$$V(r) = \frac{1}{2\pi i}\oint dz \left(\frac{\hbar z}{2}\right)\frac{d}{dz}\ln\left|\tilde{\tilde{A}}(z)\right|, \tag{14}$$

where the integration is performed in the complex frequency plane around a contour enclosing the positive part of the real frequency axis. This equation is obtained from the generalized argument principle[2]. With a standard procedure[2] one arrives at an integral along the imaginary frequency axis, $V(r) = (\hbar/2\pi)\int_0^\infty d\omega \ln\left|\tilde{\tilde{A}}(i\omega)\right|$, and the resulting interaction potential is

$$V(r) = \frac{\hbar}{2\pi}\int_0^\infty d\omega \ln\left\{\left[1 - \tilde{\alpha}_1(i\omega)\tilde{\alpha}_2(i\omega)e^{-2\omega r/c}\left[-1 + 2(\omega r/c) - (\omega r/c)^2\right]^2\right] \\ \times \left[1 - \tilde{\alpha}_1(i\omega)\tilde{\alpha}_2(i\omega)e^{-2\omega r/c}\left[-1 + \tfrac{1}{2}(\omega r/c) - \tfrac{1}{2}(\omega r/c)^2\right]^2\right]^2\right\}. \tag{15}$$

Expanding the logarithm, assuming that the polarizabilities are very small, gives

$$V(r) = -\frac{\hbar}{2\pi}\int_0^\infty d\omega\, \tilde{\alpha}_1(i\omega)\tilde{\alpha}_2(i\omega)e^{-2\omega r/c}\left[3 - 6(\omega r/c) + (17/2)(\omega r/c)^2 \\ -(9/2)(\omega r/c)^3 + (3/2)(\omega r/c)^4\right]. \tag{16}$$

Two separation limits emerge, the van der Waals limit for small separations, and the Casimir limit for large. In the van der Waals limit we have

$$V(r) \approx -(3\hbar/2\pi)\int_0^\infty d\omega\, \tilde{\alpha}_1(i\omega)\tilde{\alpha}_2(i\omega); \quad r \ll c/\omega_0, \tag{17}$$

where $\omega_0$ is some characteristic frequency above which the polarizabilities are negligible. Note that the interaction potential lacks an *r*-dependence in this range. If we assume that the characteristic frequencies are the same for the two composite particles and that the so-called London[4] approximation, $\tilde{\alpha}(\omega) = \tilde{\alpha}(0)/\left[1-(\omega/\omega_0)^2\right]$, may be used for the polarizabilities the potential in the van der Waals limit becomes $V(r) = -(3/8)\tilde{\alpha}_1(0)\tilde{\alpha}_2(0)\hbar\omega_0$.

In the Casimir limit we have

$$V(r) = -(25\hbar c/32\pi r)\tilde{\alpha}_1(0)\tilde{\alpha}_2(0); \quad r \gg c/\omega_0. \tag{18}$$

Here, the potential has the same *r*-dependence as the gravitational potential. Through the identification $\tilde{\alpha}_i(0) = m_i\sqrt{\gamma 32\pi/25\hbar c}$, we reproduce the gravitational interaction, $V(r) = -\gamma m_1 m_2/r$, where $\gamma$ is the gravitational constant. The polarizability is unit less and very small. For a universal mass unit the value is $\tilde{\alpha}_u(0) = m_u\sqrt{\gamma 32\pi/25\hbar c} \approx 1.529 \times 10^{-19}$. Thus our assumption of small polarizabilities holds for all atom nuclei.

Since we have no indication that the gravitation between two nucleons saturates before close proximity there should be no van der Waals limit. A rough estimate of the transition point between the two regions is found by equating the Casimir and van der Waals potentials, using the London approximation for the latter. Doing so one finds that $r_c \approx (25c/12\pi\omega_0)$. Demanding that this is smaller that the closest possible distance between the nucleons, ~$10^{-15}$ m, gives that the characteristic energy $\hbar\omega_0 \geq (25\hbar c/12\pi r_c) \approx 0.1\ GeV$.

In summary we have found that if one makes the assumption that the nucleons are made up by particles having harmonic oscillator interactions the resulting dispersion force between these composite particles has, in the Casimir limit, the same distance dependence as the gravitational force. If the dispersion interaction is the true origin of gravitation the mass of a particle is basically the static polarizability of the particle, $m = \sqrt{25\hbar c/\gamma 32\pi}\tilde{\alpha}(0)$. Quantizing the new field, we have introduced here, leads to a new type of gravitons. If we assume that these gravitons are in thermal equilibrium at some, probably very low, temperature the interaction will have a limited range; the interaction will drastically drop at a distance determined by this temperature. This will affect the interpretation of the accelerated expansion[5] of the Universe.

**Acknowledgements** This research was sponsored by EU, within the EU-project NANOCASE, VR, Linné Centre LiLi-NFM, and Carl Trygger's foundation.